\documentclass[letterpaper]{article} 
\usepackage{aaai2026}  
\usepackage{times}  
\usepackage{helvet}  
\usepackage{courier}  
\usepackage[hyphens]{url}  
\usepackage{graphicx} 
\usepackage{natbib}  
\usepackage{caption} 
\usepackage{algorithm}
\usepackage{algorithmic}
\usepackage{amsmath}

\usepackage{booktabs}
\usepackage{multirow}
\usepackage{newfloat}
\usepackage{listings}
\DeclareCaptionStyle{ruled}{labelfont=normalfont,labelsep=colon,strut=off} 
\floatstyle{ruled}
\newfloat{listing}{tb}{lst}{}
\floatname{listing}{Listing}
\title{BugSweeper: Function-Level Detection of Smart Contract Vulnerabilities Using Graph Neural Networks}
\author{
    Uisang Lee, Changhoon Chung, Junmo Lee, Soo-Mook Moon\thanks{Corresponding author.}
}
\affiliations{
    Seoul National University\\

    \{uisang.lee, ch.chung, junmo.lee, smoon\}@snu.ac.kr
}

\begin{document}

\maketitle
\begin{abstract}
The rapid growth of Ethereum has made it more important to quickly and accurately detect smart contract vulnerabilities. While machine learning–based methods have shown some promise, many still rely on \mbox{rule-based} preprocessing designed by domain experts. \mbox{Rule-based} preprocessing methods often discard crucial context from the source code, potentially causing certain vulnerabilities to be overlooked and limiting adaptability to newly emerging threats. We introduce BugSweeper, an end-to-end deep learning framework that detects vulnerabilities directly from the source code without manual engineering. BugSweeper represents each Solidity function as a Function-Level Abstract Syntax Graph (FLAG), a novel graph that combines its Abstract Syntax Tree (AST) with enriched control-flow and data-flow semantics. Then, our two-stage Graph Neural Network (GNN) analyzes these graphs. The first-stage GNN filters noise from the syntax graphs, while the second-stage GNN conducts high-level reasoning to detect diverse vulnerabilities. Extensive experiments on real-world contracts show that BugSweeper significantly outperforms all state-of-the-art detection methods. By removing the need for handcrafted rules, our approach offers a robust, automated, and scalable solution for securing smart contracts without any dependence on security experts. 
\end{abstract}

\begin{links}
\link{Extended version}{https://arxiv.org/abs/2512.09385}
\end{links}

\section{Introduction}

Blockchain is a decentralized and distributed ledger that enables open participation without third-party intermediaries and has attracted significant interest from academia and industry \cite{krichen:blockchain}. Ethereum extended blockchain capabilities by introducing smart contracts, which are digital agreements written in Solidity code that automatically execute transactions \cite{buterin2013ethereum}. However, these smart-contract programs introduce potential security vulnerabilities that can be exploited by malicious attackers. According to an empirical study \cite{10.1145/3377811.3380364} that analyzed 47,368 smart contracts, many vulnerabilities, such as reentrancy and unchecked low-level calls, were reported. In particular, the DAO attack \cite{TheDAO} exploited a reentrancy vulnerability to steal 3.6 million Ether (valued at \$60 million at the time).
Furthermore, smart contracts cannot be modified once deployed, unlike traditional software. Fixing a deployed contract typically requires deleting the original and redeploying an updated version, which can be both inconvenient and costly. For these reasons, it is crucial to thoroughly verify the security of smart contracts before deployment.

A variety of code-analysis techniques have been proposed to detect vulnerabilities, including static analysis \cite{tikhomirov2018smartcheck,feist2019slither,wang2024efficiently}, symbolic execution \cite{luu2016making,mueller2017mythril,mossberg2019manticore}, and dynamic execution \cite{jiang2018contractfuzzer,10.1109/ASE51524.2021.9678888,liu2018reguard}. However, these conventional methods heavily rely on manually crafted expert rules, making them ineffective against the rapid emergence of new vulnerabilities that bypass predefined patterns.

To overcome these drawbacks, researchers have increasingly leveraged deep learning models for smart contract vulnerability detection. For instance, Peculiar \cite{wu2021peculiar} and ReVulDL \cite{zhang2022reentrancy} utilize GraphCodeBERT \cite{guographcodebert}, while TMP \cite{zhuang2021smart} and AME \cite{liu2021smart} apply Graph Neural Networks (GNNs). These deep learning-based approaches can reduce analysis time and minimize reliance on expert-crafted rules. However, in a given smart contract, only a small fraction of the code is typically involved in a vulnerability. This observation motivates extracting vulnerability-specific code fragments for training vulnerability detection models. Existing deep learning methods incorporate preprocessing steps for extraction. However, these still depend on rigid, \mbox{rule-based} heuristics, resulting in several limitations:

\begin{itemize}
    \item \textbf{Restricted Scope}: They overlook vulnerabilities that are not captured by predefined rules. For example, novel variations of reentrancy attacks that deviate from established heuristics may remain undetected.
    
    \item \textbf{Poor Generalization}: Deep learning models relying on narrow, \mbox{rule-based} preprocessing cannot identify other vulnerability types (e.g., \textit{unchecked low-level calls}, \textit{arithmetic errors}) that do not fit existing patterns.

    \item \textbf{Information Loss}: If preprocessing rules are inaccurately specified, crucial details in the original code may be lost. 
\end{itemize}

In this paper, we introduce BugSweeper, a novel graph-based framework for detecting vulnerabilities in smart contracts. BugSweeper employs a GNN trained on our proposed Function-Level Abstract Syntax Graph (FLAG) representation. It effectively identifies various vulnerability types in a fully automated and data-driven manner without relying on predefined expert rules, thereby overcoming the limitations of previous \mbox{rule-based} preprocessing methods. 

\begin{figure}[t]
  \centering
  \includegraphics[width=\linewidth]{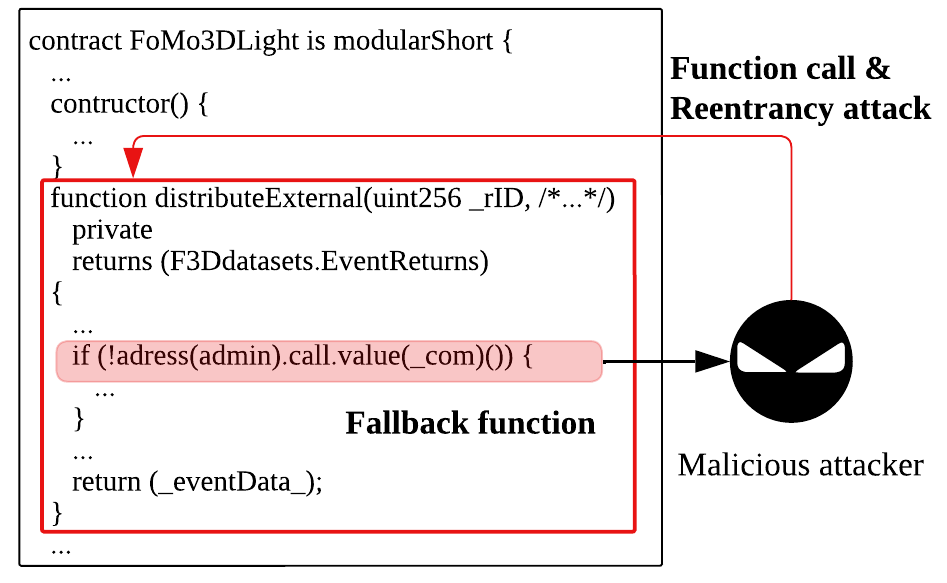}
  \caption{An example scenario demonstrating a reentrancy attack.}
\end{figure}

Figure 1 provides an example of a reentrancy attack. In this scenario, the \texttt{distributeExternal} function uses \texttt{call.value()} to send Ether to an external address, which in turn triggers the fallback function of a malicious contract. The fallback function—a default function that executes when no other function matches, or when Ether is received—then repeatedly calls \texttt{distributeExternal}, causing it to re-enter and execute again before the previous invocation completes. Such cases illustrate that many security vulnerabilities in smart contracts arise from unsafe inter-function interactions. Inspired by this observation and motivated by the limitations of current \mbox{rule-based} preprocessing methods, we propose analyzing smart contracts at the function level for more precise vulnerability detection.

While analyzing code at the function level provides precise details, it neglects important connections between functions and still leads to redundant information. To solve this, we first convert Solidity code into abstract syntax trees (ASTs) and then divide these trees into separate \mbox{function-level} subtrees. We enrich these subtrees by adding edges representing function calls and variable references, creating FLAGs. Additionally, we introduce a parameter called \textit{coverage} to control the number of inter-function connections included.

However, higher coverage settings introduce extra noise, complicating the learning process. To address this, we develop a two-stage GNN architecture. The first stage, a Code Graph Neural Network (CGNN), summarizes AST node details at the function level to mitigate noise and produces pooled function graphs. Then, the second-stage GNN analyzes these pooled graphs to capture meaningful relationships between functions.

Our extensive experiments demonstrate that BugSweeper significantly outperforms both traditional static analysis tools and existing deep learning models. Overall, it detects diverse vulnerability classes in an end-to-end, data-driven manner without relying on predefined expert rules.

This paper makes the following contributions:
\begin{itemize}
\item \textbf{A Novel Data-Driven, Multi-Class Vulnerability Detection Framework}: We introduce BugSweeper, the first unified and rule-free framework that detects multiple vulnerability classes directly from code structure. Experiments conducted on three distinct vulnerability types—(i) Reentrancy, (ii) Unchecked Low-Level Calls, and (iii) Time Manipulation—demonstrate the effectiveness of our \mbox{function-level}, data-driven approach that generalizes to diverse vulnerabilities.

\item \textbf{Function-Level Representation with Two-Stage GNN Architecture}: We introduce FLAG, a specialized graph representation that captures the rich semantic and structural features of Solidity code relevant to security analysis. Furthermore, our two-stage GNN \mbox{architecture}—comprising a CGNN and a second-stage GNN—effectively reduces AST-level noise and captures meaningful relationships among pooled function graphs.

\end{itemize}

\begin{figure*}[t]
  \centering
  \includegraphics[width=0.9\linewidth]{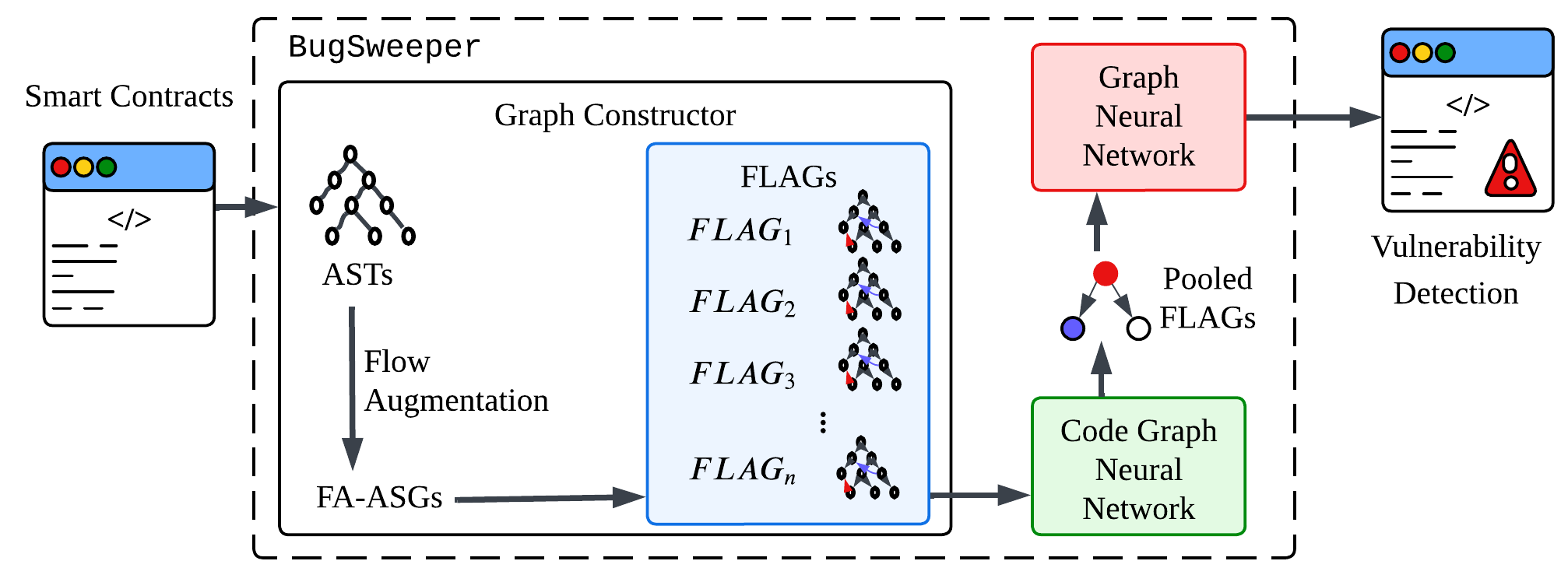}
  \caption{The overall architecture of BugSweeper. BugSweeper parses each contract into an AST, augments it with control-flow and data-flow edges to form Function-Level Abstract Syntax Graphs (FLAGs), and then applies a two-stage GNN: a Code Graph Neural Network, followed by a second-stage GNN.}
\end{figure*}

\section{Related Work}
\subsubsection{Traditional Smart Contract Analysis Methods.} Traditional methods, which rely on static analysis and symbolic execution, have been widely used to identify vulnerabilities in smart contracts. Specifically, static analysis-based methods detect vulnerability patterns by analyzing code, data flow, and control flow using predefined rules, without executing the program. Static analysis-based tools for smart contract vulnerability detection include SmartCheck \cite{tikhomirov2018smartcheck} and Slither \cite{feist2019slither}. Symbolic execution-based methods identify vulnerabilities by tracking program execution paths and collecting symbolic values for conditions along those paths. Representative tools include Oyente \cite{luu2016making}, Osiris \cite{torres2018osiris}, Manticore \cite{mossberg2019manticore}, and Mythril \cite{mueller2017mythril}. Recent methods, such as Slise \cite{wang2024efficiently}, employ a hybrid approach that combines static analysis with symbolic execution.
\subsubsection{Deep Learning-Based Smart Contract Analysis Methods.} Various deep learning-based methods have been developed for vulnerability detection. TMP \cite{zhuang2021smart} converts contracts into symbolic graphs, refines them by removing less important nodes, and normalizes the graph to apply Graph Neural Networks (GNNs). Furthermore, AME \cite{liu2021smart} combines expert patterns with TMP. In contrast, Peculiar \cite{wu2021peculiar} and ReVulDL \cite{zhang2022reentrancy} focus on a contract's data flow, targeting critical variables within a data-flow graph to analyze reentrancy vulnerabilities using the GraphCodeBERT model \cite{guographcodebert}. However, current deep learning approaches have not fully escaped the rigidity of earlier methods. Many still depend on \mbox{rule-based} components, such as the explicit expert patterns in AME or the heuristic-based graph simplification in TMP. This reliance on heuristics and specialization fundamentally limits their applicability to various vulnerabilities.

\section{BugSweeper}
In this section, we introduce BugSweeper, a GNN-based framework for detecting smart contract vulnerabilities, and describe its architecture in three parts: the Graph Constructor, the Code Graph Neural Network, and the Second-Stage Graph Neural Network.

\subsection{Graph Constructor}
Previous research showed that AST-based approaches perform well on many code-related tasks \cite{zhang2019novel}, and FA-AST \cite{wang2020detecting} demonstrated that adding data-flow and control-flow details to an AST helps the model outperform a basic AST approach. Building on this, we use an abstract syntax graph that integrates data and control flow.
\begin{figure*}[t]
  \centering
  \includegraphics[width=0.99\linewidth]{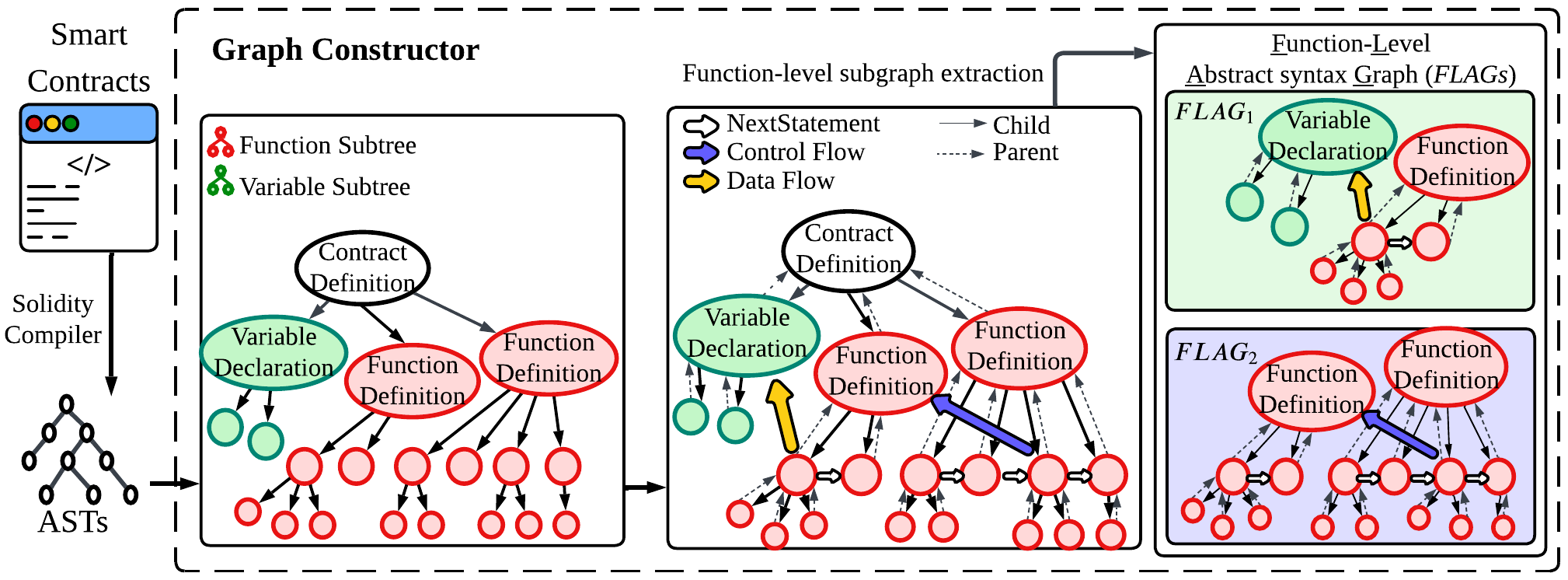}
  \caption{An overview of FLAG (Function-Level Abstract Syntax Graph) construction. The Graph Constructor enriches the AST with explicit control-flow (blue) and data-flow (yellow) edges. We then extract \mbox{function-level} subgraphs, called FLAGs, which represent target functions while preserving key connections to related code. The example uses coverage = 2, so each FLAG also includes directly connected functions and variables.}
\end{figure*}

Figure 3 illustrates how we construct FLAGs. First, we use the Solidity compiler \textit{solc} as a parser to convert the source code into an AST. To enable \mbox{function-level} extraction, we mark the nodes in subtrees whose root is a \texttt{FunctionDefinition} or \texttt{VariableDeclaration}. A basic AST fails to capture critical program execution details, so we augment it with supplementary edges to form a flow-augmented abstract syntax graph (FA-ASG). Specifically, we introduce control-flow and data-flow edges to create a richer semantic representation. Our FA-ASG consists of three distinct edge categories: the original structural edges from the AST, along with the newly added control-flow and data-flow edges, as detailed below.
\begin{itemize}
\item Basic edges capture the fundamental structure of the AST. Specifically, \textit{Child} edges link parent nodes to their child nodes, while \textit{Parent} edges connect child nodes back to parent nodes, thereby improving overall node connectivity \cite{allamanis2018learning}.
\item Data-flow edges illustrate how data moves through a program. Specifically, \textit{ReferencedDeclaration} edges indicate the use of previously defined variables or functions. \textit{FunctionReturnParameter} edges directly link a \texttt{FunctionDefinition} node to the data it returns. \textit{SuperFunction} edges represent function-overriding relationships. \textit{Assignment} edges capture the process of assigning data to variables that have already been declared.
\item Control-flow edges show how the execution sequence flows in \texttt{IfStatement}, \texttt{LoopStatement}, and \texttt{NextStatement} nodes. For an \texttt{IfStatement}, \textit{CondTrue} indicates the path when the condition is true, while \textit{CondFalse} shows the path when it is false. A \texttt{LoopStatement} includes \textit{WhileExecution} and \textit{ForExecution} edges that point to the condition, and \textit{WhileNext} and \textit{ForNext} edges leading to the statements executed once the condition is met. \texttt{NextStatement} edges represent the sequential ordering of statements.
\end{itemize}
Next, the contract is divided into multiple subgraphs, where each subgraph represents a single function. To model the dependencies between these functions, we introduce a hyperparameter termed coverage. This parameter controls the depth of neighborhood expansion when constructing each function's graph, determining the extent to which subgraphs of called functions or referenced variables are included. This concept is illustrated in Figure 4.

For instance, a coverage of 1 includes only the subgraph of the target function itself. A coverage of 2 expands the graph to include the subgraphs of all directly called functions and variables (the 1-hop neighborhood). At a coverage of 3, the graph further incorporates the subgraphs of functions and variables called by that second level (the 2-hop neighborhood). Through this iterative expansion, we generate enriched FLAGs that serve as our final units for analysis.

\begin{figure*}[t]
  \centering
  \includegraphics[width=0.98\linewidth]{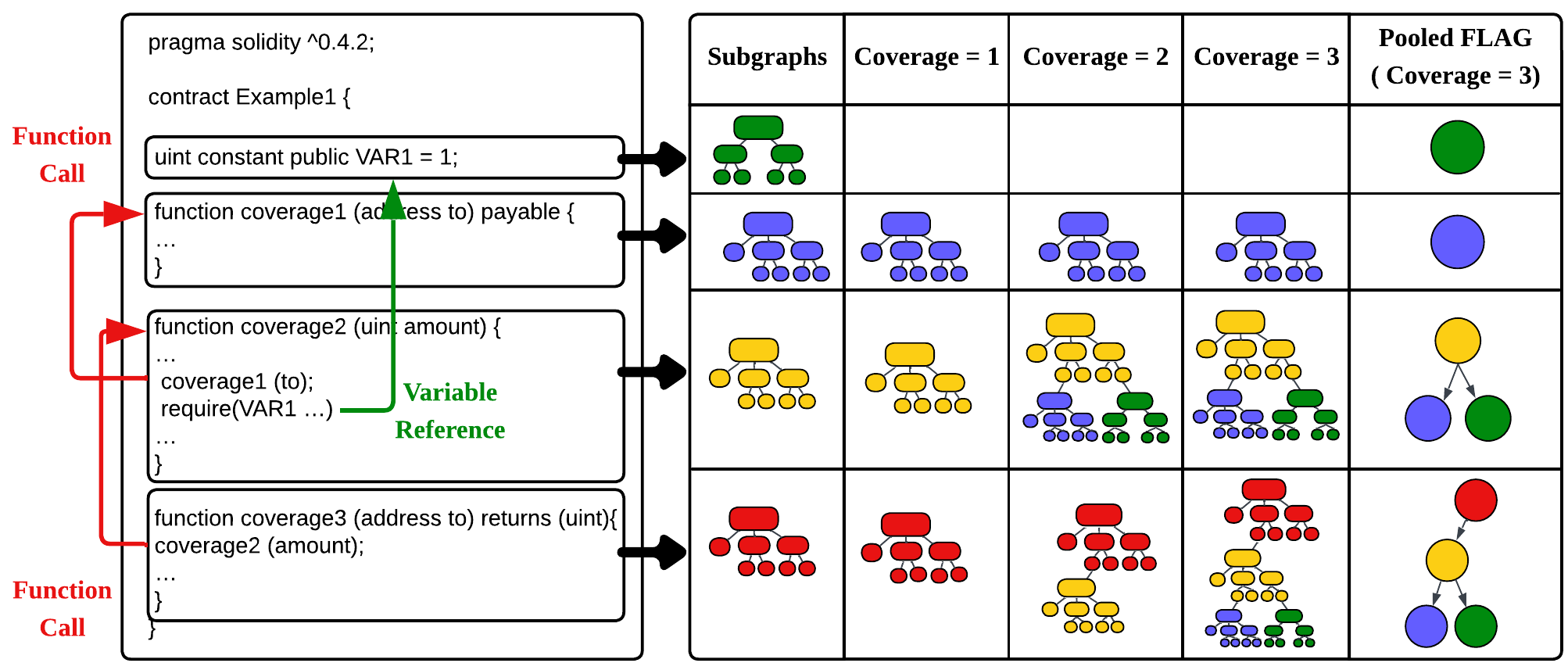}
  \caption{Solidity code is parsed into an AST and decomposed into function- and variable-level subgraphs. Higher coverage incorporates more inter-function connections. The graph for \texttt{coverage1} stays the same across coverage levels, as it does not involve any function calls or variable references. Colors in the figure are solely for distinguishing the subgraphs.}
\end{figure*}

\subsection{Code Graph Neural Network (CGNN)}
In this section, we explain how CGNN converts each FLAG into a Pooled FLAG, reducing unnecessary graph noise through pooling techniques. Each node in a FLAG carries textual attributes (e.g., \texttt{FunctionDefinition}). To convert the textual node attributes into vector representations suitable for neural network input, we apply the BPE tokenizer \cite{sennrich2015neural}.

After tokenizing the node attributes, we apply GraphSAGE \cite{10.5555/3294771.3294869}, an efficient, inductive GNN, to generate detailed node embeddings. GraphSAGE updates each node representation by aggregating features from its neighboring nodes using the following message-passing mechanism:

\begin{equation}
    h^{k}_v \xleftarrow{} AGG^{k}\Big(\Big\{\big(h_u^{(k-1)}:u\in\mathcal{N}(v)\big)\Big\}\Big) \oplus h^{k-1}_v,
\end{equation}
\begin{equation}
    h^{k}_v \xleftarrow{}\sigma(\mathbf{W}^{k}\cdot h^{k}_v),
\end{equation}
where $k$ is the layer number for message-passing depth $K$, $h^{0}_v$ is the initial node feature encoded via a BPE tokenizer, and $\mathcal{N}(v)$ refers to the neighborhood function, which maps each node $v$ to its set of neighboring nodes in the graph. AGG aggregates feature vectors of neighboring nodes, and $\oplus$ represents the concatenation operation that merges the aggregated message with $h_{v}^{k-1}$. Table 1 shows the overall architecture of CGNN.

For graph-level vulnerability classification, individual node embeddings must be aggregated into a single representation. While global mean and max pooling are commonly used in many GNN applications, they are suboptimal for our FLAGs. These naive methods fail to distinguish between core functional nodes and auxiliary nodes introduced by external function calls or variable references, often yielding noisy and uninformative graph embeddings.

Several advanced graph pooling methods have been proposed, such as TopKPool \cite{gao2019graph}, SAGPool \cite{lee2019self}, and ASAPool \cite{ranjan2020asap}. TopKPool and SAGPool select and retain nodes by scoring them based on node features and attention mechanisms, respectively. On the other hand, ASAPool groups nodes into clusters based on local subgraph scores. Although effective, TopKPool and SAGPool often face scalability issues on large graphs and can unintentionally discard critical structural information. While ASAPool addresses some of these issues, its clustering process is computationally expensive, limiting its practicality for large-scale applications.

\begin{table}[t]
  \setlength{\tabcolsep}{4pt}
  \renewcommand{\arraystretch}{0.8}
  \centering\small{
  \begin{tabular}{lccc}
    \toprule
    \textbf{Layer}&\textbf{Input Size}&\textbf{Output Size}&\textbf{Activation Function}\\
    \midrule
    SAGEConv& 512& 1024& ReLU\\
    SAGEConv& 1024& 1024& ReLU\\
    SAGEConv& 1024& 1024& -\\ 
  \bottomrule
\end{tabular}}
\caption{CGNN architecture}
\end{table}

Rather than relying on computationally expensive clustering or learned node-scoring mechanisms, we propose a deterministic and efficient semantic pooling method called Code Graph Pool (CGPool). It groups nodes based on their syntactic roles in the source code—for example, merging all nodes that belong to the same function declaration or variable definition into a single supernode.

This design preserves key high-level relationships by reconnecting these supernodes according to their original control-flow and data-flow links. The resulting structure, which we term the Pooled FLAG, is a compact yet faithful abstraction of the program's logic (as shown in Figure 4). CGPool effectively integrates the benefits of previous pooling strategies. Like ASAPool, it maintains important hierarchical structures but without the computational cost. Similar to TopKPool and SAGPool, CGPool reduces graph complexity without the risk of information loss from aggressive node selection.

\begin{table}[t]
  \setlength{\tabcolsep}{1.9pt}
  \renewcommand{\arraystretch}{0.8}
  \centering\small{
  \begin{tabular}{lcccc}
    \toprule
    \textbf{Layer} & \textbf{Input Size} & \textbf{Output Size} & \textbf{Activation Function} & \textbf{Heads} \\
    \midrule
    GATConv   & 1024 & 1024 & ReLU & 4 \\
    GATConv   & 1024 & 1024 & ReLU & 1 \\
    GATConv   & 1024 & 1024 & ReLU & 1 \\
    Linear1   & 1024 & 1024 & ReLU & – \\
    Linear2   & 1024 & 1024 & ReLU & – \\
    Linear3   & 1024 & $C$  & –    & – \\
    \bottomrule
  \end{tabular}}
  \caption{Second-stage GNN and classifier architecture. $C$ refers to the number of vulnerability classes.}
  \label{tab:GNNs}
\end{table}

\subsection{Second-Stage Graph Neural Network}

After the pooling step, we obtain a simplified FLAG where each node represents a meaningful group of elements from the original code, such as functions or variables. This compact structure significantly reduces the number of nodes while preserving essential structural information. Because of this reduction, we can efficiently apply a GNN to capture higher-level relationships among the supernodes.

We use a Graph Attention Network (GAT) \cite{velivckovic2017graph} to process the Pooled FLAG. GAT assigns dynamic weights to edges based on their importance, allowing the model to focus more on critical connections—such as those between function nodes or between a function and its related variables. This produces rich node embeddings that encode both local and contextual information. After the GAT layers, we apply a global readout function to summarize the entire graph into a single vector. This vector reflects the vulnerability patterns identified in the contract and serves as the input to the classifier.

The classifier consists of three linear layers: the first two use ReLU activation, and the final layer applies a \mbox{softmax} function to produce class probabilities. It is trained using numerous examples of source code paired with their corresponding labels to predict whether a given contract contains vulnerable functions. Although we demonstrate BugSweeper's effectiveness across three vulnerability types, our model can be extended to detect a broader range of vulnerabilities. Table 2 summarizes the architecture of the second-stage GNN and classifier used in BugSweeper.

\section{Experiments}
\subsection{Datasets and Experimental Setup}
\subsubsection{Dataset.} 

We use the AME dataset \cite{liu2021smart} to compare the performance of various methods for detecting reentrancy vulnerabilities. Other methods were either limited to detecting only reentrancy vulnerabilities or did not specify their preprocessing steps, making it challenging to reproduce results for additional vulnerability types. The AME dataset contains 1,224 smart contracts, which we divide into 719 contracts for training and 505 contracts for testing, following the original experimental setup for accurate performance comparisons.

Additionally, we use the SmartBugs Wild dataset \cite{10.1145/3377811.3380364}, consisting of 47,398 Solidity files with a total of 203,716 contracts, to evaluate our model’s performance on multi-class vulnerability detection tasks and conduct ablation studies. The SmartBugs Wild dataset labels contracts as vulnerable if three or more heuristic methods agree on the presence of a vulnerability, significantly reducing false positives and ensuring reliable ground truth labels. The dataset is structured based on this consensus approach as follows:
\begin{itemize}
    \item \textbf{{Reentrancy}}: Reentrancy happens when a contract function calls an external contract before updating its state, which can allow attackers to re-enter the function multiple times. A contract is labeled with this vulnerability only if confirmed by four or more tools, ensuring high confidence for this critical bug.
    \item \textbf{{Unchecked Low-Level Calls}}: This category includes vulnerabilities related to functions like $\mathtt{call()}$ or $\mathtt{send()}$ that fail without proper checks. For this type, a vulnerability is labeled as positive if detected by a consensus of three or more tools.
    \item \textbf{{Time Manipulation}}: In Solidity, $\mathtt{block.timestamp}$ can be manipulated by miners, posing security risks. These vulnerabilities are labeled as positive if detected by a consensus of at least three tools.
\end{itemize}

\subsubsection{Experimental Setup.}  
We evaluate performance using macro-averaged Precision, Recall, and F1-score \cite{wu2021peculiar}. Macro-averaging calculates metrics for each class individually and then averages them, making it suitable for vulnerability detection tasks, which typically involve highly imbalanced datasets. This ensures equal contribution of each class to the overall evaluation. We train all models using the Adam optimizer \cite{kingma2014adam} with a learning rate of 1e-4 and weight decay of 1e-5. We train for up to 500 epochs with a batch size of 64. For robustness, each experiment was repeated four times with different random seeds, and all table entries report the mean over these runs. In most comparisons, the performance improvements are statistically significant (paired $t$-test, $p \leq 0.05$). The mean $\pm$ standard deviation results and full paired $t$-test statistics are given in the Appendix.

\subsection{Reentrancy Detection Results}

We evaluate our proposed method, BugSweeper, against four baseline vulnerability detection tools—Slither \cite{feist2019slither}, SmartCheck \cite{tikhomirov2018smartcheck}, Mythril \cite{mueller2017mythril}, and Slise \cite{wang2024efficiently}—as well as four deep learning-based techniques, TMP \cite{zhuang2021smart}, AME \cite{liu2021smart}, Peculiar \cite{wu2021peculiar}, and ReVulDL \cite{zhang2022reentrancy}.

Table 3 presents the results of our experiments, highlighting several important findings. First, traditional methods (Slither, SmartCheck, Mythril, Slise) exhibited poor performance in detecting reentrancy vulnerabilities, as evidenced by their low recall scores. Second, deep learning-based methods achieved significantly better performance overall. Notably, our \mbox{function-level} approach, BugSweeper, outperformed all other models on the AME dataset. BugSweeper demonstrated an improvement in precision (up to 99.87\%) and achieved the highest overall F1-score (98.57\%), outperforming the closest competitor by approximately 3.1\%. These results indicate that BugSweeper significantly reduces both false positives and false negatives compared to existing approaches.

Overall, our findings confirm the effectiveness of BugSweeper’s \mbox{function-level}, rule-independent approach for detecting reentrancy vulnerabilities. By constructing FLAGs that integrate both syntax structure and control- and data-flow information, BugSweeper adapts to diverse vulnerability patterns—resulting in greater flexibility, robustness, and accuracy in smart contract security analysis.

\begin{table}[t]
\setlength{\tabcolsep}{7pt}
\renewcommand{\arraystretch}{0.9}
\centering\small{
\begin{tabular}{lccc} 
\toprule
\multicolumn{1}{l}{\multirow{2}{*}{\textbf{Approach}}} & \multicolumn{3}{c}{\textbf{Reentrancy}}                       \\ 
\cmidrule{2-4}
\multicolumn{1}{c}{}                          & Precision (\%)          & Recall (\%)        & F1 (\%)               \\ 
\midrule
Slither     & 94.74          & 34.62          & 50.70           \\
SmartCheck      & 88.57          & 55.36          & 68.13          \\
Mythril   & 84.48          & 47.12          & 60.49           \\ 
Slise   & 95.56          & 45.26          & 61.43           \\ 
\midrule
TMP     &  87.06 & 83.74  & 85.36  \\
AME     &  95.45 & 95.38  & 95.42  \\
Peculiar     & 92.58  & 94.40  & 93.48  \\
ReVulDL     & 92.95   & 94.62  & 93.74  \\
BugSweeper (ours)    & \textbf{99.87}   & \textbf{97.35}  & \textbf{98.57}  \\
\bottomrule
\end{tabular}}
\caption{Performance comparison of various approaches for detecting reentrancy vulnerabilities. The first four methods represent traditional \mbox{rule-based} detection tools, while the remaining approaches are deep learning–based. }
\end{table}

\begin{table*}[t]
\setlength{\tabcolsep}{1.5pt}
\renewcommand{\arraystretch}{0.8}
\centering\small{
\begin{tabular}{@{}llcccccccccc@{}}
\toprule
\multirow{2}{*}{\textbf{Architecture}} & \multicolumn{2}{l}{\multirow{2}{*}{\textbf{Model}}} & \multicolumn{3}{c}{\textbf{Reentrancy}}          & \multicolumn{3}{c}{\textbf{Unchecked Low-Level Calls}} & \multicolumn{3}{c}{\textbf{Time Manipulation}}   \\ \cmidrule(l){4-12} 
 & \multicolumn{2}{c}{}   & Precision (\%)  & Recall (\%)     & F1 (\%)   & Precision (\%)    & Recall (\%)       & F1 (\%)     & Precision (\%)  & Recall (\%)     & F1 (\%)   \\ \midrule
\multirow{2}{*}{%
  \begin{tabular}[c]{@{}l@{}}
    BugSweeper \\
    (single-stage GNN)
  \end{tabular}
} & \multicolumn{2}{l}{GAT} & 90.28 & 78.23 & 84.77 & 81.86 & 68.33  & 71.43 & 90.07 & 63.88 & 74.72 \\
                               & \multicolumn{2}{l}{SAGE} & 87.77  & 78.97 & 83.11  & 83.32  & 54.97  & 66.20  & 88.69  & 64.68  & 74.77  \\ \cmidrule(r){0-11}
\multirow{4}{*}{BugSweeper}    & \multicolumn{2}{l}{GAT + SAGE}  & 82.24  & 82.82 & 82.46  & 83.20  & 66.67 & 74.02  & 78.83  & 61.89 & 69.33  \\
                & \multicolumn{2}{l}{GAT + GAT} & 86.42  & 85.38 & 85.66 & 89.01  & 60.81  & 72.16 & 81.14 & 63.88  & 72.50  \\
                & \multicolumn{2}{l}{SAGE + SAGE} & 87.59 & 90.25 & 88.89 & 83.41 & 69.01 & 75.38 & 84.58  & 67.46  & 75.06           \\
& \multicolumn{2}{l}{SAGE + GAT}  & 90.91 & \textbf{92.31} & \textbf{91.61}  & 85.70 & \textbf{75.44}  & \textbf{80.15} & 89.27  & 69.44 & \textbf{79.63} \\ \bottomrule
\end{tabular}}
\caption{Performance analysis of different GNN configurations within the BugSweeper framework on the SmartBugs Wild dataset; the analysis was conducted with a coverage of 4. The results confirm that our two-stage architecture significantly outperforms the single-stage baseline, with the SAGE + GAT model achieving the highest F1-scores. Bolded results are the best and statistically significant ($p \leq 0.05$, paired $t$-test).}
\end{table*}

\subsection{Multi-class Vulnerability Detection Results}

To evaluate BugSweeper's capability in detecting multiple vulnerability types, we conducted multi-class classification experiments focusing on three categories: (i) Reentrancy, (ii) Unchecked Low-Level Calls, and (iii) Time Manipulation. Additionally, we investigated the impact of different GNN configurations on detection performance.

Table 4 illustrates that our proposed two-stage BugSweeper models outperform the single-stage baseline models across all vulnerability types. Specifically, the two-stage approach significantly improves performance, effectively capturing complex vulnerability patterns.

Among the tested configurations, the combination of GraphSAGE in the first stage and GAT in the second stage (SAGE + GAT) achieved the highest performance, obtaining an F1-score of 91.61\% for reentrancy, 80.15\% for unchecked low-level calls, and 79.63\% for time manipulation. These results demonstrate how GraphSAGE and GAT play distinct but synergistic roles. GraphSAGE excels at aggregating information across large, detail-rich graphs, while GAT’s attention mechanism sharpens focus on the most critical features in the pooled, higher-level representation.

However, we observed comparatively lower detection performance for unchecked low-level calls and time manipulation vulnerabilities relative to reentrancy. This difference is due to data imbalance, as noted in \cite{10.1145/3377811.3380364}. The smaller number of examples for these vulnerabilities makes it more challenging for the model to learn effective representations, resulting in somewhat lower performance across our evaluation metrics.

\begin{table}[t]
\centering\resizebox{\linewidth}{!}{
\begin{tabular}{@{}llccc@{}}
\toprule
\multicolumn{1}{l}{\multirow{2}{*}{\textbf{Model}}} & \multirow{2}{*}{\textbf{Pool}} & \multicolumn{3}{c}{\textbf{Multi-class Vulnerability Detection}} \\ \cmidrule(l){3-5}
\multicolumn{1}{c}{}    &\multicolumn{1}{c}{}                             & Precision (\%)       & Recall (\%)         & F1 (\%)       \\ 
\midrule
\multirow{3}{*}{\shortstack[l]{BugSweeper \\ (CGNN only)
}}
   & TopKPool          & 87.74 & 72.48 & 78.71 \\
   & SAGPool           & 83.91 & 69.29 & 75.13 \\
   & ASAPool           & 89.85 & 74.02 & 80.32 \\ \midrule
\multirow{4}{*}{\shortstack[l]{BugSweeper \\ (SAGE + GAT)}} 
   & TopKPool          & 84.77 & 68.86 & 75.29 \\
   & SAGPool           & 88.59 & 69.55 & 77.10 \\
   & ASAPool           & 87.37 & 78.60 & 82.41 \\ \cmidrule(l){2-5}
   & CGPool (ours)     & 91.27 & \textbf{84.21} & \textbf{87.32} \\ 
\bottomrule
\end{tabular}
}
\caption{Multi‑class vulnerability detection performance for different pooling methods on BugSweeper variants. BugSweeper (CGNN only) denotes a single-stage CGNN with GraphSAGE. Bolded results are the best and statistically significant.}
\end{table}

\subsection{Ablation Studies}
\subsubsection{Component Study.} 
To validate the effectiveness of BugSweeper’s architecture, we conducted a detailed ablation study, with results shown in Table 4. The goal of this experiment was to understand how each component contributes to overall performance.

First, we created a baseline model representing a simpler, single-stage GNN. This baseline does not include our proposed CGPool or the two-stage GNN structure. Instead, it uses a standard GNN architecture (GAT or GraphSAGE) followed by a global pooling layer to produce the final graph representation.

Next, we evaluated our complete two-stage BugSweeper architecture, testing several combinations of GNN models across the two stages. For instance, a configuration labeled GAT + SAGE indicates that we use GAT for the first stage (CGNN) and GraphSAGE in the second stage to analyze the pooled graph representation.

Our experiments provide two key insights. First, the two-stage architecture consistently outperforms the single-stage baseline, confirming that the second-stage GNN effectively refines and enhances the representations produced by the first stage. Second, the combination of GraphSAGE in the first stage, followed by GAT in the second stage, consistently achieves the highest overall F1-score. This result demonstrates a complementary synergy: GraphSAGE initially captures a broad structural context, and GAT’s attention mechanism precisely identifies critical features within the reduced, second-stage graph.

\subsubsection{Pooling Method.}
To evaluate the effectiveness of our domain-specific CGPool, we replace CGPool in BugSweeper with three widely used graph pooling methods—TopKPool \cite{gao2019graph}, SAGPool \cite{lee2019self}, and ASAPool \cite{ranjan2020asap}—and evaluate their performance on multi-class vulnerability detection (Table 5). TopKPool scores nodes using learned projection vectors and retains only the highest-scoring subset, while SAGPool extends this idea by incorporating neighbor information via graph convolutions. ASAPool, in contrast, hierarchically clusters local subgraphs to capture rich structural patterns. For comparison, we evaluate a simplified variant that removes the second-stage GNN, leaving a single CGNN. In this single-stage setting, we apply one of these pooling methods and then use global mean pooling on its output to obtain the final graph representation.

The results in Table 5 demonstrate two key findings. First, substituting CGPool with any of the pooling methods in the BugSweeper architecture yields lower F1‐scores (75.13–82.41\%) than our CGPool baseline (87.32\%), confirming that CGPool’s AST-aware semantic clustering is better suited to code analysis. 

Second, our experiments validate the effectiveness of the two-stage GNN architecture for this task. Pooling methods demonstrate improved performance with a second GNN layer, except for TopKPool, indicating that this second stage can refine the features from the initial pooling step to enhance detection performance.

In summary, CGPool not only achieves the highest F1-score but also maintains balanced precision (91.27\%) and recall (84.21\%) compared to other methods, showing its domain effectiveness for \mbox{function-level} vulnerability detection.
\section{Conclusion}

In this paper, we present BugSweeper, a function-level framework for detecting vulnerabilities in smart contracts. BugSweeper has two core components: a Graph Constructor that builds Function-Level Abstract Syntax Graphs (FLAGs) from the source code, and a two-stage \mbox{Graph Neural Network (GNN)} that first mitigates noise and then performs high-level reasoning over the extracted graphs. In addition, our domain-specific pooling method, Code Graph Pool (CGPool), effectively reduces information loss during graph abstraction. Experimental results demonstrate that BugSweeper not only outperforms existing approaches in detecting reentrancy vulnerabilities but also remains effective across multiple vulnerability categories. Ultimately, our framework enhances the accuracy and robustness of vulnerability detection, contributing to more secure and reliable smart-contract ecosystems. 

\section{Acknowledgments}
This work was supported in part by the National Research Foundation of Korea (NRF) grant funded by Korean Government [Ministry of Science and ICT (MSIT)] under Grant RS-2023-00208245, 30\%; in part by the Institute of Information and Communications Technology Planning and Evaluation (IITP) grant funded by Korean Government (MSIT) under Grant 2021-0-00180, 40\%; in part by the Information Technology Research Center (ITRC) support Program Supervised by IITP under Grant IITP-2021-0-01835, 10\%; and in part by IITP under the Graduate School of Artificial Intelligence Semiconductor Grant IITP-2025-RS-2023-00256081, 10\%; and in part by Hyundai Motor Chung Mong-Koo Foundation, 10\%.

\bibliography{aaai2026}

@article{buterin2013ethereum,
  title={Ethereum white paper},
  author={Buterin, Vitalik and others},
  journal={GitHub repository},
  volume={1},
  pages={22--23},
  year={2013}
}

@inproceedings{10.1145/3377811.3380364,
author = {Durieux, Thomas and Ferreira, Jo\~{a}o F. and Abreu, Rui and Cruz, Pedro},
title = {Empirical review of automated analysis tools on 47,587 Ethereum smart contracts},
year = {2020},
isbn = {9781450371216},
publisher = {Association for Computing Machinery},
address = {New York, NY, USA},
url = {https://doi.org/10.1145/3377811.3380364},
doi = {10.1145/3377811.3380364},
booktitle = {Proceedings of the ACM/IEEE 42nd International Conference on Software Engineering},
pages = {530–541},
numpages = {12},
location = {Seoul, South Korea},
series = {ICSE '20}
}

@online{TheDAO,
  author = {Phil Daian},
  title = {Analysis of the DAO exploit},
  year = 2016,
  url = {https://hackingdistributed.com/2016/06/18/analysis-of-the-dao-exploit/},
  urldate = {2016-06-18}
}

@misc{mueller2017mythril,
  title={Mythril-Reversing and bug hunting framework for the Ethereum blockchain},
  author={Mueller, B},
  journal={2022-01-16]. https://pypi. org/project/mythril/0.8. 2},
  year={2017}
}

@inproceedings{luu2016making,
author = {Luu, Loi and Chu, Duc-Hiep and Olickel, Hrishi and Saxena, Prateek and Hobor, Aquinas},
title = {Making Smart Contracts Smarter},
year = {2016},
isbn = {9781450341394},
publisher = {Association for Computing Machinery},
address = {New York, NY, USA},
url = {https://doi.org/10.1145/2976749.2978309},
doi = {10.1145/2976749.2978309},
booktitle = {Proceedings of the 2016 ACM SIGSAC Conference on Computer and Communications Security},
pages = {254–269},
numpages = {16},
location = {Vienna, Austria},
series = {CCS '16}
}

@inproceedings{torres2018osiris,
author = {Torres, Christof Ferreira and Sch\"{u}tte, Julian and State, Radu},
title = {Osiris: Hunting for Integer Bugs in Ethereum Smart Contracts},
year = {2018},
isbn = {9781450365697},
publisher = {Association for Computing Machinery},
address = {New York, NY, USA},
url = {https://doi.org/10.1145/3274694.3274737},
doi = {10.1145/3274694.3274737},
booktitle = {Proceedings of the 34th Annual Computer Security Applications Conference},
pages = {664–676},
numpages = {13},
location = {San Juan, PR, USA},
series = {ACSAC '18}
}

@inproceedings{mossberg2019manticore,
author = {Mossberg, Mark and Manzano, Felipe and Hennenfent, Eric and Groce, Alex and Grieco, Gustavo and Feist, Josselin and Brunson, Trent and Dinaburg, Artem},
title = {Manticore: a user-friendly symbolic execution framework for binaries and smart contracts},
year = {2020},
isbn = {9781728125084},
publisher = {IEEE Press},
url = {https://doi.org/10.1109/ASE.2019.00133},
doi = {10.1109/ASE.2019.00133},
booktitle = {Proceedings of the 34th IEEE/ACM International Conference on Automated Software Engineering},
pages = {1186–1189},
numpages = {4},
location = {San Diego, California},
series = {ASE '19}
}

@inproceedings{tikhomirov2018smartcheck,
author = {Tikhomirov, Sergei and Voskresenskaya, Ekaterina and Ivanitskiy, Ivan and Takhaviev, Ramil and Marchenko, Evgeny and Alexandrov, Yaroslav},
title = {SmartCheck: static analysis of ethereum smart contracts},
year = {2018},
isbn = {9781450357265},
publisher = {Association for Computing Machinery},
address = {New York, NY, USA},
url = {https://doi.org/10.1145/3194113.3194115},
doi = {10.1145/3194113.3194115},
booktitle = {Proceedings of the 1st International Workshop on Emerging Trends in Software Engineering for Blockchain},
pages = {9–16},
numpages = {8},
location = {Gothenburg, Sweden},
series = {WETSEB '18}
}

@inproceedings{feist2019slither,
author = {Feist, Josselin and Greico, Gustavo and Groce, Alex},
title = {Slither: a static analysis framework for smart contracts},
year = {2019},
publisher = {IEEE Press},
url = {https://doi.org/10.1109/WETSEB.2019.00008},
doi = {10.1109/WETSEB.2019.00008},
booktitle = {Proceedings of the 2nd International Workshop on Emerging Trends in Software Engineering for Blockchain},
pages = {8–15},
numpages = {8},
location = {Montreal, Quebec, Canada},
series = {WETSEB '19}
}

@inproceedings{jiang2018contractfuzzer,
author = {Jiang, Bo and Liu, Ye and Chan, W. K.},
title = {ContractFuzzer: fuzzing smart contracts for vulnerability detection},
year = {2018},
isbn = {9781450359375},
publisher = {Association for Computing Machinery},
address = {New York, NY, USA},
url = {https://doi.org/10.1145/3238147.3238177},
doi = {10.1145/3238147.3238177},
booktitle = {Proceedings of the 33rd ACM/IEEE International Conference on Automated Software Engineering},
pages = {259–269},
numpages = {11},
location = {Montpellier, France},
series = {ASE '18}
}

@inproceedings{10.1109/ASE51524.2021.9678888,
author = {Choi, Jaeseung and Kim, Doyeon and Kim, Soomin and Grieco, Gustavo and Groce, Alex and Cha, Sang Kil},
title = {SMARTIAN: enhancing smart contract fuzzing with static and dynamic data-flow analyses},
year = {2022},
isbn = {9781665403375},
publisher = {IEEE Press},
url = {https://doi.org/10.1109/ASE51524.2021.9678888},
doi = {10.1109/ASE51524.2021.9678888},
booktitle = {Proceedings of the 36th IEEE/ACM International Conference on Automated Software Engineering},
pages = {227–239},
numpages = {13},
location = {Melbourne, Australia},
series = {ASE '21}
}

@inproceedings{liu2018reguard,
author = {Liu, Chao and Liu, Han and Cao, Zhao and Chen, Zhong and Chen, Bangdao and Roscoe, Bill},
title = {ReGuard: finding reentrancy bugs in smart contracts},
year = {2018},
isbn = {9781450356633},
publisher = {Association for Computing Machinery},
address = {New York, NY, USA},
url = {https://doi.org/10.1145/3183440.3183495},
doi = {10.1145/3183440.3183495},
booktitle = {Proceedings of the 40th International Conference on Software Engineering: Companion Proceedings},
pages = {65–68},
numpages = {4},
location = {Gothenburg, Sweden},
series = {ICSE '18}
}

@inproceedings{zhuang2021smart,
author = {Zhuang, Yuan and Liu, Zhenguang and Qian, Peng and Liu, Qi and Wang, Xiang and He, Qinming},
title = {Smart contract vulnerability detection using graph neural networks},
year = {2021},
isbn = {9780999241165},
booktitle = {Proceedings of the Twenty-Ninth International Joint Conference on Artificial Intelligence},
articleno = {454},
numpages = {8},
location = {Yokohama, Yokohama, Japan},
series = {IJCAI'20}
}

@inproceedings{wu2021peculiar,
  title={Peculiar: Smart contract vulnerability detection based on crucial data flow graph and pre-training techniques},
  author={Wu, Hongjun and Zhang, Zhuo and Wang, Shangwen and Lei, Yan and Lin, Bo and Qin, Yihao and Zhang, Haoyu and Mao, Xiaoguang},
  booktitle={2021 IEEE 32nd International Symposium on Software Reliability Engineering (ISSRE)},
  pages={378--389},
  year={2021},
  organization={IEEE}
}

@inproceedings{zhang2022reentrancy,
author = {Zhang, Zhuo and Lei, Yan and Yan, Meng and Yu, Yue and Chen, Jiachi and Wang, Shangwen and Mao, Xiaoguang},
title = {Reentrancy Vulnerability Detection and Localization: A Deep Learning Based Two-phase Approach},
year = {2023},
isbn = {9781450394758},
publisher = {Association for Computing Machinery},
address = {New York, NY, USA},
url = {https://doi.org/10.1145/3551349.3560428},
doi = {10.1145/3551349.3560428},
booktitle = {Proceedings of the 37th IEEE/ACM International Conference on Automated Software Engineering},
articleno = {83},
numpages = {13},
location = {Rochester, MI, USA},
series = {ASE '22}
}

@inproceedings{velivckovic2017graph,
  title={Graph Attention Networks},
  author={Veli{\v{c}}kovi{\'c}, Petar and Cucurull, Guillem and Casanova, Arantxa and Romero, Adriana and Li{\`o}, Pietro and Bengio, Yoshua},
  booktitle={International Conference on Learning Representations},
  year={2018}
}

@inproceedings{10.5555/3294771.3294869,
author = {Hamilton, William L. and Ying, Rex and Leskovec, Jure},
title = {Inductive representation learning on large graphs},
year = {2017},
isbn = {9781510860964},
publisher = {Curran Associates Inc.},
address = {Red Hook, NY, USA},
booktitle = {Proceedings of the 31st International Conference on Neural Information Processing Systems},
pages = {1025–1035},
numpages = {11},
location = {Long Beach, California, USA},
series = {NIPS'17}
}

@INPROCEEDINGS{wang2020detecting,
author = { Wang, Wenhan and Li, Ge and Ma, Bo and Xia, Xin and Jin, Zhi },
booktitle = { 2020 IEEE 27th International Conference on Software Analysis, Evolution and Reengineering (SANER) },
title = {{ Detecting Code Clones with Graph Neural Network and Flow-Augmented Abstract Syntax Tree }},
year = {2020},
volume = {},
ISSN = {1534-5351},
pages = {261-271},
doi = {10.1109/SANER48275.2020.9054857},
url = {https://doi.ieeecomputersociety.org/10.1109/SANER48275.2020.9054857},
publisher = {IEEE Computer Society},
address = {Los Alamitos, CA, USA},
month =Feb}

@Article{krichen:blockchain,
AUTHOR = {Krichen, Moez and Ammi, Meryem and Mihoub, Alaeddine and Almutiq, Mutiq},
TITLE = {Blockchain for Modern Applications: A Survey},
JOURNAL = {Sensors},
VOLUME = {22},
YEAR = {2022},
NUMBER = {14},
ARTICLE-NUMBER = {5274},
URL = {https://www.mdpi.com/1424-8220/22/14/5274},
PubMedID = {35890953},
ISSN = {1424-8220},
DOI = {10.3390/s22145274}
}

@inproceedings{allamanis2018learning,
  title={Learning to Represent Programs with Graphs},
  author={Allamanis, Miltiadis and Brockschmidt, Marc and Khademi, Mahmoud},
  booktitle={International Conference on Learning Representations},
  year={2018}
}

@inproceedings{liu2021smart,
  title={Smart Contract Vulnerability Detection: From Pure Neural Network to Interpretable Graph Feature and Expert Pattern Fusion},
  author={Liu, Zhenguang and Qian, Peng and Wang, Xiang and Zhu, Lei and He, Qinming and Ji, Shouling},
   booktitle={Proceedings of the Twenty-Ninth International Conference on International Joint Conferences on Artificial Intelligence},
  pages={2751--2759},
  year={2021}
}

@inproceedings{zhang2019novel,
  title={A novel neural source code representation based on abstract syntax tree},
  author={Zhang, Jian and Wang, Xu and Zhang, Hongyu and Sun, Hailong and Wang, Kaixuan and Liu, Xudong},
  booktitle={2019 IEEE/ACM 41st International Conference on Software Engineering (ICSE)},
  pages={783--794},
  year={2019},
  organization={IEEE}
}

@inproceedings{sennrich2015neural,
    title = "Neural Machine Translation of Rare Words with Subword Units",
    author = "Sennrich, Rico  and
      Haddow, Barry  and
      Birch, Alexandra",
    editor = "Erk, Katrin  and
      Smith, Noah A.",
    booktitle = "Proceedings of the 54th Annual Meeting of the Association for Computational Linguistics (Volume 1: Long Papers)",
    month = aug,
    year = "2016",
    address = "Berlin, Germany",
    publisher = "Association for Computational Linguistics",
    url = "https://aclanthology.org/P16-1162/",
    doi = "10.18653/v1/P16-1162",
    pages = "1715--1725"
}

@inproceedings{guographcodebert,
  title={GraphCodeBERT: Pre-training code representations with data flow},
  author={Guo, Daya and Ren, Shuo and Lu, Shuai and Feng, Zhangyin and Tang, Duyu and Liu, Shujie and Zhou, Long and Duan, Nan and Svyatkovskiy, Alexey and Fu, Shengyu and others},
  booktitle={International Conference on Learning Representations},
  year={2021}
}

@InProceedings{gao2019graph,
  title = 	 {Graph U-Nets},
  author =       {Gao, Hongyang and Ji, Shuiwang},
  booktitle = 	 {Proceedings of the 36th International Conference on Machine Learning},
  pages = 	 {2083--2092},
  year = 	 {2019},
  editor = 	 {Chaudhuri, Kamalika and Salakhutdinov, Ruslan},
  volume = 	 {97},
  series = 	 {Proceedings of Machine Learning Research},
  month = 	 {09--15 Jun},
  publisher =    {PMLR},
  url = 	 {https://proceedings.mlr.press/v97/gao19a.html}
}

@InProceedings{lee2019self,
  title = 	 {Self-Attention Graph Pooling},
  author =       {Lee, Junhyun and Lee, Inyeop and Kang, Jaewoo},
  booktitle = 	 {Proceedings of the 36th International Conference on Machine Learning},
  pages = 	 {3734--3743},
  year = 	 {2019},
  editor = 	 {Chaudhuri, Kamalika and Salakhutdinov, Ruslan},
  volume = 	 {97},
  series = 	 {Proceedings of Machine Learning Research},
  month = 	 {09--15 Jun},
  publisher =    {PMLR},
  url = 	 {https://proceedings.mlr.press/v97/lee19c.html}
}

@article{wang2024efficiently,
author = {Wang, Zexu and Chen, Jiachi and Wang, Yanlin and Zhang, Yu and Zhang, Weizhe and Zheng, Zibin},
title = {Efficiently Detecting Reentrancy Vulnerabilities in Complex Smart Contracts},
year = {2024},
issue_date = {July 2024},
publisher = {Association for Computing Machinery},
address = {New York, NY, USA},
volume = {1},
number = {FSE},
url = {https://doi.org/10.1145/3643734},
doi = {10.1145/3643734},
journal = {Proc. ACM Softw. Eng.},
month = jul,
articleno = {8},
numpages = {21}
}

@inproceedings{ranjan2020asap,
  title={Asap: Adaptive structure aware pooling for learning hierarchical graph representations},
  author={Ranjan, Ekagra and Sanyal, Soumya and Talukdar, Partha},
  booktitle={Proceedings of the AAAI conference on artificial intelligence},
  volume={34},
  number={04},
  pages={5470--5477},
  year={2020}
}

@article{kingma2014adam,
  title={Adam: A method for stochastic optimization},
  author={Kingma, Diederik P and Ba, Jimmy},
  journal={arXiv preprint arXiv:1412.6980},
  year={2014}
}

\cleardoublepage
\appendix

\section{A. Experimental Details}

\subsubsection{Training Setup.} We implemented BugSweeper using PyTorch and PyTorch Geometric libraries. All experiments were conducted on a single NVIDIA RTX 4090 GPU with 24GB memory. Each experiment was repeated four times with different random seeds, and the reported performance is the average with standard deviation.

\subsubsection{Hyperparameter Configuration.} The following hyperparameters were used in our experiments:

\begin{table}[h]
\centering
\small
\begin{tabular}{|l|c|}
\hline
\textbf{Hyperparameter} & \textbf{Search Range} \\
\hline
Learning Rate & [1e-5, 1e-3] \\
Weight Decay & [1e-5, 1e-3] \\
Batch Size & [32, 64] \\
Dropout & [0.3, 0.5] \\
Coverage (FLAG) & [1, 5] \\
\hline
\end{tabular}
\caption{Search ranges of hyperparameters explored during model selection.}
\end{table}

For all main experiments, we used the following hyperparameter settings: learning rate of 1e-4, weight decay of 1e-5, batch size of 64. The dropout rate was set to 0.3 for the classifier and 0.5 for the Graph Neural Networks. We fixed the random seed at 42 and set the coverage parameter to 4.

\subsubsection{Dataset.} 
For the reentrancy detection experiments, we used the AME dataset, which consists of 1,224 labeled smart contracts. Following the original experiment configuration, we split the dataset into 719 contracts for training and 505 for testing to ensure consistent and fair comparison with prior work. 
For model selection, we further set aside 10\% of the training contracts as a validation split, sampled at random.

At the function level, the resulting splits are as follows:
\begin{itemize}
    \item Train: The dataset contains 4,305 non-reentrant functions, 215 reentrant functions, and a total of 4,520 functions.

    \item Test: It comprises 3,159 non-reentrant functions and 151 reentrant functions, for a total of 3,310 functions.
\end{itemize}

For multi-class vulnerability detection, we utilized a subset of the SmartBugs Wild dataset. For per-class validation, we constructed a balanced split with an approximate 2:1:1 ratio for training, validation, and testing, respectively. The final dataset includes 462 contracts labeled with reentrancy, 88 with unchecked low-level calls, and 49 with time manipulation vulnerabilities. 

At the function level, the splits contain the following number of labeled functions:

\begin{itemize}
    \item Train: 10,454 Safe, 363 Reentrancy, 95 Unchecked Low-Level Calls, and 189 Time-Manipulation functions, for a total of 11,101 functions.
    \item Validation: 5,271 Safe, 191 Reentrancy, 59 Unchecked Low-Level Calls, and 81 Time-Manipulation functions, for a total of 5,602 functions.
    \item Test: 4,919 Safe, 130 Reentrancy, 57 Unchecked Low-Level Calls, and 84 Time-Manipulation functions, for a total of 5,190 functions.
\end{itemize}

\section{B. Coverage}

To evaluate the effect of the coverage parameter, we varied the parameter from~1 to~5.
As shown in Table 7, the size of the preprocessed dataset grows almost linearly from 355 MB at coverage 1 to 742 MB at coverage~4, and then increases only slightly (762 MB at coverage~5).
While higher coverage provides richer context that may reveal additional vulnerabilities, it also introduces greater noise.
Balancing detection accuracy, we use \textbf{coverage~=~4} for all main experiments.
\begin{table}[h]
\centering
\small
\begin{tabular}{|c|c|c|}
\hline
\textbf{Coverage} & \textbf{Preprocessed Data Size (MB)} & \textbf{Test F1 (\%)} \\
\hline
1 & 355 (Baseline) & 86.57 \\
2 & 560 ($1.57\times$) & 85.05 \\
3 & 691 ($1.94\times$)& 87.61 \\
4 & 742 ($2.09\times$)& 87.73 \\
5 & 762 ($2.14\times$)& 86.92 \\
\hline
\end{tabular}
\caption{
Effect of the coverage parameter on preprocessed data size and test F1-score.
Coverage controls the neighborhood expansion during graph construction.
Values in parentheses indicate the relative data-size increase with respect to the baseline
(coverage\,=\,1).
}
\end{table}

\section{C. Algorithms}

\paragraph{Function-Level Graph Construction.}
We construct a function-level abstract code graph (FLAG) for each contract by recursively expanding each function’s local subgraph to include referenced functions and variables, up to the coverage parameter. Edge types, such as control flow and variable reference, are added during graph construction. The final output is a set of enriched subgraphs for the second-stage graph neural network. See Algorithm 1 for detailed FLAG construction, and Algorithms 2-3 for the two-stage GNN inference procedure.

\begin{algorithm}[h]
\caption{Graph Constructor}
\label{alg:flag_construction}
\textbf{Input}: Solidity source code $C$\\
\textbf{Parameter}: coverage parameter $k$\\
\textbf{Output}: set of function‐level graphs $\mathcal{G}$, FLAG

\begin{algorithmic}[1]
\STATE $AST \leftarrow \mathrm{Parse}(C)$ using \texttt{solc} compiler
\STATE $G \leftarrow \mathrm{BuildGraph}(AST)$ using \texttt{NetworkX}
\STATE $\mathcal{G} \leftarrow \{\}$
\FOR{each function subtree $f$ in $G$}
    \STATE Initialize $S \leftarrow \{\text{subtree rooted at }f\}$
    \FOR{$d = 1$ \TO $k$}
        \STATE Expand $S$ by adding subgraphs of all functions and variables referenced by nodes in $S$
    \ENDFOR
    \STATE $G_f \leftarrow \mathrm{MergeSubgraphs}(S)$
            \STATE Add data‐flow edges and control‐flow edges to $G_f$
    \STATE $\mathcal{G}.\mathrm{append}(G_f)$
\ENDFOR
\STATE \textbf{return} $\mathcal{G}$
\end{algorithmic}
\end{algorithm}

\paragraph{Code Graph Neural Network.}
Our code graph neural network first encodes each node’s code tokens and aggregates nodes into function or variable subgraphs using the pool mapping derived from the FLAG construction.

\begin{algorithm}[h]
\caption{Code Graph Neural Network}
\label{alg:cgcnn}
\textbf{Input}: Node features $X$, edge indices $E$, pool mapping $P$\\
\textbf{Parameter}:  Number of GraphSAGE layers $L_1$\\
\textbf{Output}: Pooled node embeddings $H$ and pooled adjacency $E'$

\begin{algorithmic}[1]
\STATE $Z \leftarrow \mathrm{Embed}(X)$  \hfill // Token embedding
\STATE $H \leftarrow \mathrm{MeanPool}(Z, P)$  \hfill // Pool per function/variable subgraph
\FOR{each GraphSAGE layer $\ell = 1 \dots L_1$}
\STATE $H \leftarrow \mathrm{GraphSAGE}_\ell(H, E)$
\STATE $H \leftarrow \mathrm{ReLU}(H)$
\STATE $H \leftarrow \mathrm{Dropout}(H, 0.5)$
\ENDFOR
\STATE $E', B' \leftarrow \mathrm{PoolAdj}(E, P)$  \hfill // Pooled adjacency and batch
\STATE \textbf{return} $H, E', B'$
\end{algorithmic}
\end{algorithm}

\paragraph{Second-Stage Graph Neural Network.}
In this stage, the pooled embeddings and adjacency from stage 1 are further analyzed using multiple GAT layers, followed by a global readout and a classifier to produce final vulnerability logits. 
\begin{algorithm}[h]
\caption{Second Stage Graph Neural Network}
\label{alg:second_stage_gnn}
\textbf{Input}: Pooled embeddings $H$, pooled adjacency $E'$, batch assignment $B'$\\
\textbf{Parameter}: Number of GAT layers $L_2$\\
\textbf{Output}: Prediction logits $Y$

\begin{algorithmic}[1]
\FOR{each GAT layer $m = 1 \dots L_2$}
\STATE $H \leftarrow \mathrm{GAT}_m(H, E')$
\STATE $H \leftarrow \mathrm{ReLU}(H)$
\STATE $H \leftarrow \mathrm{Dropout}(H, 0.5)$
\ENDFOR
\STATE $h_G \leftarrow \mathrm{GlobalMeanPool}(H, B')$  \hfill // Graph-level readout
\STATE $Y \leftarrow \mathrm{Classifier}(h_G)$  \hfill // Final classification
\STATE \textbf{return} $Y$
\end{algorithmic}
\end{algorithm}

\section{D. Statistical Test Results}

To assess the statistical significance of performance improvements, we conducted paired $t$-tests across all model comparisons. We report the p-values for key evaluation metrics (Precision, Recall, and F1-score) across three different experimental settings:

\begin{itemize}
    \item \textbf{Table 8}: BugSweeper vs. baseline vulnerability detection models on the Reentrancy task.
    
    \item \textbf{Table 9}: Comparisons of five GNN variants (GAT, SAGE, GAT+SAGE, GAT+GAT, SAGE+SAGE) against the SAGE+GAT configuration within BugSweeper, evaluated under both the single-stage and two-stage architectures across three vulnerability types—Reentrancy, Unchecked Low-Level Calls (denoted as \textbf{Unchecked.}), and Time Manipulation (denoted as \textbf{Time Manip.}). The table reports precision, recall, and F1-scores, along with paired t-test p-values computed by comparing each variant to the \textbf{SAGE+GAT} configuration for each metric.

    \item \textbf{Table 10}: Comparisons of our pooling method \textbf{CGPool} against alternative pooling strategies (TopKPool, SAGPool, and ASAPool) for multi-class vulnerability detection, evaluated on BugSweeper with and without the second-stage GNN (SAGE backbone). The table reports precision, recall, and F1-scores, along with paired t-test p-values computed by comparing each pooling strategy to \textbf{CGPool (ours)} for each metric.

\end{itemize}

Tables 8–10 collectively demonstrate that BugSweeper consistently outperforms prior baseline detectors and strong ablations across multiple vulnerability types and evaluation settings. In particular, the two-stage design and our architectural choices—such as the SAGE+GAT backbone configuration and the proposed CGPool—yield higher precision, recall, and F1-scores in most cases. Moreover, the accompanying paired t-test results indicate that these gains are statistically significant for the majority of comparisons, underscoring that the improvements are not incidental but attributable to the proposed design choices.

\begin{table}[h]
\centering{
\begin{tabular}{lcccc}
\toprule
\textbf{Metric} & \textbf{TMP} & \textbf{AME} & \textbf{Peculiar} & \textbf{ReVulDL} \\
\midrule
Prec. & $<$ 0.001* & $<$ 0.001* & $<$ 0.001* & $<$ 0.001* \\
Rec.    & $<$ 0.001* & 0.026* & 0.001* & 0.001* \\
F1  & $<$ 0.001* & 0.001* & $<$ 0.001* & $<$ 0.001* \\
\bottomrule
\end{tabular}}
\caption{Paired t-test p-values comparing BugSweeper with baseline models on the Reentrancy task (Table 3). * denotes significance at 5\% level.}
\end{table}

\begin{table*}[h]
\centering
\small
\begin{tabular}{@{}lllcccccc@{}}
\toprule
\multirow{2}{*}{\textbf{Architecture}} &
\multirow{2}{*}{\textbf{Model}} &
\multirow{2}{*}{\textbf{Vulnerability}} &
\multicolumn{3}{c}{\textbf{Performance}} &
\multicolumn{3}{c}{\textbf{Paired t-test p-values}} \\
\cmidrule(l){4-6}\cmidrule(l){7-9}
& & &
Precision(\%) & Recall(\%) & F1(\%) &
$p_{\text{Prec}}$ & $p_{\text{Rec}}$ & $p_{\text{F1}}$ \\
\midrule

\multirow{6}{*}{\shortstack[l]{BugSweeper \\ (single-stage GNN)}}
& \multirow{3}{*}{GAT}
& Reentrancy   & 90.28 ($\pm$ 0.94) & 78.23 ($\pm$ 1.60) & 84.77 ($\pm$ 1.26) & 0.786 & 0.013* & 0.050* \\
& & Unchecked.  & 81.86 ($\pm$ 3.17) & 68.33 ($\pm$ 7.89) & 71.43 ($\pm$ 4.86) & 0.234 & 0.310  & 0.074  \\
& & Time Manip. & 90.07 ($\pm$ 3.84) & 63.88 ($\pm$ 0.68) & 74.72 ($\pm$ 1.09) & 0.771 & 0.034* & 0.031* \\
\cmidrule(l){2-9}
& \multirow{3}{*}{SAGE}
& Reentrancy   & 87.77 ($\pm$ 2.31) & 78.97 ($\pm$ 3.20) & 83.11 ($\pm$ 2.19) & 0.290 & 0.007* & 0.007* \\
& & Unchecked.  & 83.32 ($\pm$ 4.85) & 54.97 ($\pm$ 2.02) & 66.20 ($\pm$ 2.49) & 0.506 & 0.001* & 0.011* \\
& & Time Manip. & 88.69 ($\pm$ 3.78) & 64.68 ($\pm$ 1.82) & 74.77 ($\pm$ 1.87) & 0.822 & 0.184  & 0.027* \\

\midrule

\multirow{12}{*}{BugSweeper}
& \multirow{3}{*}{GAT + SAGE}
& Reentrancy   & 82.24 ($\pm$ 6.20) & 82.82 ($\pm$ 1.77) & 82.46 ($\pm$ 3.51) & 0.058 & 0.026* & 0.024* \\
& & Unchecked.  & 83.20 ($\pm$ 1.38) & 66.67 ($\pm$ 1.75) & 74.02 ($\pm$ 1.60) & 0.437 & 0.038* & 0.006* \\
& & Time Manip. & 78.83 ($\pm$ 1.75) & 61.89 ($\pm$ 1.19) & 69.33 ($\pm$ 0.43) & 0.012* & 0.034* & 0.001* \\
\cmidrule(l){2-9}
& \multirow{3}{*}{GAT + GAT}
& Reentrancy   & 86.42 ($\pm$ 6.93) & 85.38 ($\pm$ 1.17) & 85.66 ($\pm$ 3.22) & 0.292 & 0.034* & 0.050* \\
& & Unchecked.  & 89.01 ($\pm$ 2.11) & 60.81 ($\pm$ 4.41) & 72.16 ($\pm$ 2.59) & 0.267 & 0.020* & 0.038* \\
& & Time Manip. & 81.14 ($\pm$ 4.61) & 63.88 ($\pm$ 3.63) & 72.50 ($\pm$ 1.44) & 0.189 & 0.107  & 0.030* \\
\cmidrule(l){2-9}
& \multirow{3}{*}{SAGE + SAGE}
& Reentrancy   & 87.59 ($\pm$ 1.55) & 90.25 ($\pm$ 0.88) & 88.89 ($\pm$ 0.44) & 0.994 & 0.206  & 0.108  \\
& & Unchecked.  & 83.41 ($\pm$ 4.61) & 69.01 ($\pm$ 4.04) & 75.38 ($\pm$ 1.39) & 0.620 & 0.186  & 0.002* \\
& & Time Manip. & 84.58 ($\pm$ 2.11) & 67.46 ($\pm$ 1.37) & 75.06 ($\pm$ 1.60) & 0.061 & 0.199  & 0.046* \\
\cmidrule(l){2-9}
& \multirow{3}{*}{SAGE + GAT}
& Reentrancy   & 90.91 ($\pm$ 2.81) & \textbf{92.31} ($\pm$ 1.33) & \textbf{91.61} ($\pm$ 2.05) & --  & -- & -- \\
& & Unchecked.  & 85.70 ($\pm$ 5.01) & \textbf{75.44} ($\pm$ 1.75) & \textbf{80.15} ($\pm$ 1.38) & -- & -- & -- \\
& & Time Manip. & 89.27 ($\pm$ 0.34) & 69.44 ($\pm$ 2.47) & \textbf{79.63} ($\pm$ 1.05) & -- & -- & -- \\

\bottomrule
\end{tabular}
\caption{Performance analysis of different GNN configurations within the BugSweeper framework on the SmartBugs Wild dataset (coverage=4). The last three columns report paired t-test $p$-values comparing each variant against the SAGE+GAT configuration for each metric; * denotes significance at the 5\% level ($p\leq$0.05).}
\end{table*}

\begin{table*}[t]
\centering
\begin{tabular}{@{}llcccccc@{}}
\toprule
\multicolumn{1}{l}{\multirow{2}{*}{\textbf{Model}}} &
\multirow{2}{*}{\textbf{Pool}} &
\multicolumn{3}{c}{\textbf{Multi-class Vulnerability Detection}} &
\multicolumn{3}{c}{\textbf{Paired t-test p-value}} \\
\cmidrule(l){3-5}\cmidrule(l){6-8}
 & & Precision (\%) & Recall (\%) & F1 (\%) & $p_{\text{Prec}}$ & $p_{\text{Rec}}$ & $p_{\text{F1}}$ \\
\midrule

\multirow{3}{*}{\shortstack[l]{BugSweeper \\ w/o second-stage GNN \\ (SAGE)}}
& TopKPool & 87.74 ($\pm$ 1.70) & 72.48 ($\pm$ 1.00) & 78.71 ($\pm$ 0.49) & 0.786 & 0.013* & 0.050* \\
& SAGPool  & 83.91 ($\pm$ 1.97) & 69.29 ($\pm$ 1.79) & 75.13 ($\pm$ 1.90) & 0.290 & 0.007* & 0.007* \\
& ASAPool  & 89.85 ($\pm$ 1.14) & 74.02 ($\pm$ 2.43) & 80.32 ($\pm$ 1.11) & 0.058 & 0.026* & 0.024* \\
\midrule

\multirow{4}{*}{\shortstack[l]{BugSweeper \\ (SAGE + GAT)}}
& TopKPool      & 84.77 ($\pm$ 2.81) & 68.86 ($\pm$ 4.31) & 75.29 ($\pm$ 3.51) & 0.234 & 0.310  & 0.074  \\
& SAGPool       & 88.59 ($\pm$ 3.13) & 69.55 ($\pm$ 2.21) & 77.10 ($\pm$ 2.72) & 0.506 & 0.001* & 0.011* \\
& ASAPool       & 87.37 ($\pm$ 2.42) & 78.60 ($\pm$ 2.92) & 82.41 ($\pm$ 1.05) & 0.437 & 0.038* & 0.006* \\
\cmidrule(l){2-8}
& CGPool (ours) & 91.27 ($\pm$ 1.89) & \textbf{84.21} ($\pm$ 0.82) & \textbf{87.32} ($\pm$ 0.78) & -- & -- & -- \\
\bottomrule
\end{tabular}
\caption{Multi-class vulnerability detection performance for different pooling methods on BugSweeper variants. The last three columns report paired t-test $p$-values comparing each pooling strategy against CGPool (ours) for each metric; * denotes significance at the 5\% level ($p\leq$0.05).}
\end{table*}

\end{document}